\documentclass{vgtc}                          

\ifpdf
  \pdfoutput=1\relax                   
  \usepackage{graphicx}                
  \DeclareGraphicsExtensions{.pdf,.png,.jpg,.jpeg} 
\else
  \usepackage{graphicx}                
  \DeclareGraphicsExtensions{.eps}     
\fi%

\graphicspath{{figures/}{pictures/}{images/}{./}} 

\usepackage{microtype}                 
\PassOptionsToPackage{warn}{textcomp}  
\usepackage{textcomp}                  
\usepackage{mathptmx}                  
\usepackage{times}                     
\usepackage{cite}                      
\usepackage{tabu}                      
\usepackage{booktabs}                  
\usepackage{cite}
\usepackage{enumitem}

\onlineid{0}

\vgtccategory{Research}

\title{Did You Get The Gist Of It?
Understanding How Visualization Impacts Decision-Making}

\author{Melanie Bancilhon\thanks{e-mail: mbancilhon@wustl.edu}\\ %
        \scriptsize Washington University in St. Louis %
\and Alvitta Ottley\thanks{e-mail: alvittao@wustl.edu}\\ %
    \scriptsize Washington University in St. Louis}

\teaser{
  \centering
  \includegraphics[width=0.7\linewidth]{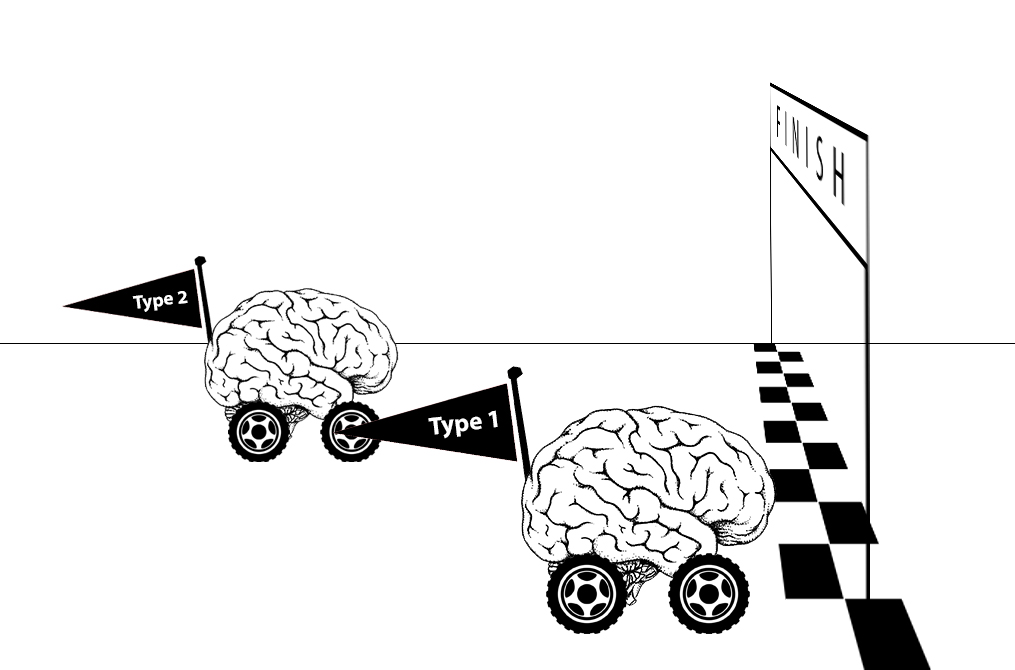}
  \caption{Illustration of Type 1 and Type 2 reasoning as conceptualized by Tversky and Kahneman. Type 1, our intuitive system, is at the forefront of decision-making processes while Type 2, our analytic system, operates secondarily.}
  \label{fig:teaser}
}

\abstract{As visualization researchers evaluate the impact of visualization design on decision-making, they often hold a one-dimensional perspective on the cognitive processes behind making a decision. Several psychological and economical researchers have shown that to make decisions, people rely on quantitative reasoning as well as gist-based intuition -- two systems that operate in parallel. In this position paper, we discuss decision theories and provide suggestions to bridge the gap between the evaluation of decision-making in visualization and psychology research. The goal is to question the limits of our knowledge and to advocate for a more nuanced understanding of decision-making with visualization.
 } 

\CCScatlist{
  \CCScatTwelve{Human-centered computing}{Visu\-al\-iza\-tion}{Visu\-al\-iza\-tion techniques}{Decision-making};
  \CCScatTwelve{Human-centered computing}{Visu\-al\-iza\-tion}{Visualization design and evaluation methods}{}
}

\begin{document}


\firstsection{Introduction}

\maketitle
We make hundreds of decisions every day, ranging from trivial to complex. Such choices could include when to leave our house to catch the bus or whether to take an umbrella when there is a high chance of rain. More complex decisions could include whether to invest in the stock market given the potential return value. Whether decisions seem complex or mundane, they are guided by two types of reasoning, commonly referred to as Type 1 and Type 2. Type 1 guides our intuition and recognition patterns while Type 2 is responsible for analytical thinking~\cite{kahneman_book}. The dual-process theory has been popularized in Daniel Kahneman's book ``Thinking Fast and Slow", where they describe Type 1 as the dominant system in charge of reasoning and judgment. 
However, this notion of two parallel systems is formalized with Fuzzy Trace Theory (FTT), where the two systems are defined as gist (high level) and verbatim (detail level) reasoning~\cite{reyna1995fuzzy}. FTT posits that people make decisions from gist-based intuition, a ``fuzzy" representation of the information extracted. This theory has challenged the prior misconception that decision-making is, in its most advanced form, rooted in pure logic.


We assert that understanding the role of Type 1 and Type 2 reasoning in decision-making with visualization is critical.
In many cases, it is increasingly common to use data visualization to support reasoning about risks and to aid sound decision-making, and its impact can be colossal~\cite{padilla2018decision}. For example, the `flatten the curve' visualization helped shape the public lexicon during the COVID-19 pandemic. While the topic of decision-making has been explored by many visualization researchers, they often focus on the binary outcome of a decision rather than the decision-making process. Their findings are often limited to the task at hand, and often contradict other research areas that evaluate visual aides using gist measures. 
In this paper, we discuss the current methods for visualization evaluation in the context of decision-making, and provide suggestions to adopt a more nuanced and holistic approach in the visualization community.

\section{Dual Process in Decision-Making}

Fuzzy Trace Theory states that people make decisions by extracting meaning from verbatim input to make a gist-based judgment. We rely on the least precise gist representation necessary to make a decision~\cite{reyna1995fuzzy}, generally referred to as ``fuzzy processing preference". Because precision is often associated with accuracy, many believe that quantitative reasoning is superior to qualitative reasoning. However, Reyna et al. showed that experts in the medical field tend to engage more in gist-based decision-making compared to novices~\cite{reyna2006}. Tversky and Kahneman made the argument that intuition is a synonym for recognition~\cite{kahneman_book}. Experts recognize familiar situations and can therefore make fast and accurate decisions even when they are complex. 

While gist reasoning has been proven to be effective, Type 1 reasoning is also more susceptible to false first impressions and framing effects. Consider the following question:
\newline
\begin{quote}\label{example-prospect}
\begin{flushleft}
A bat and ball cost \$1.10. The bat costs \$1 more than the ball. How much does the ball cost?
\end{flushleft}
\end{quote}

More than 50\% of students at Harvard, Princeton and the Massachusetts Institute of Technology routinely give the incorrect answer, insisting the ball costs 10 cents~\cite{kahneman_book}~\footnote{The correct answer to this problem is that the ball costs 5 cents and the bat costs — at a dollar more — \$1.05 for a grand total of \$1.10.}. Type 1 1 is at the forefront of cognitive processes, and it often requires significant effort to switch from Type 1 to Type 2 reasoning in order to avoid cognitive biases. 

Before the acknowledgement of the role of Type 1, many believed that Type 2 was solely in charge of decision-making operations. Expected Utility Theory, prevalent in the field of economics, posits that people make decisions rationally, using Type 2 reasoning to compute the utility of events. The recognition of dual modes of reasoning lead to the development of \textit{Prospect Theory}~\cite{kahneman2013prospect}, and revolutionized the way economists think about decision-making.

\begin{figure}[b!]
    \centering
    \includegraphics[width=.45\textwidth]{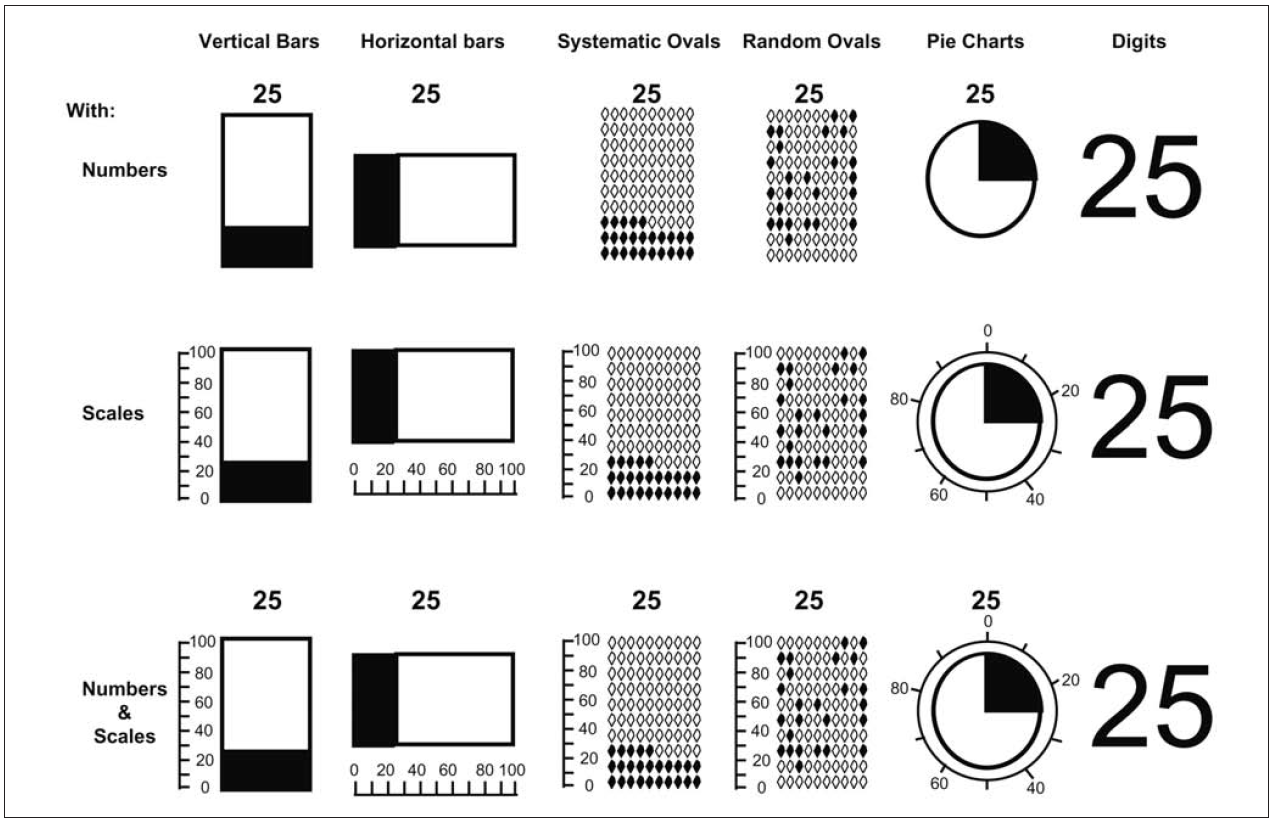}
    \caption{The 6 visualization designs used in Feldman's study ~\cite{feldman2007}. Participants were shown two percentages (in 1 format) and asked to choose i) which one was larger/smaller (gist) and ii) estimate the size of the difference (verbatim). Their response time was recorded to observe the ease of communication of visualization designs. Other measures were investigated such as the effect of background color and the presence of scales and numbers.}
    
    \label{fig:lotterysheet}
\end{figure}

\section{Visualization and Decision-Making}
Many scholars have investigated the impact of visualization on risk perception and decisions
~\cite{greis2018,fernandes2018,bisantz2011,kale2019,kay2016ish,correll2018}. A number of these studies evaluate the impact of visualization design on decisions by prompting participants for probability estimates. For example, Kay et al. evaluated visualizations in their ability to communicate the uncertainty of transit data by asking participants to estimate the likeliness of bus arrival times on a scale of 0 to 100~\cite{kay2016ish}. We believe that prompting participants for a numerical estimate compels them to use their Type 2 reasoning to understand the visualization and does not reflect how most people make decisions in real-life: by gisting. Others have used realistic simulations to observe the direct impact of visualization design on decisions. Kay et al. has observed people's ability to leave the house on time to catch the bus while minimizing waiting time at the bus stop \cite{kay2016ish}. Greis et al. has used a Facebook game “Farm Smart” to observe how participants best grow and sell crops given uncertainty weather information \cite{greis2018}. Bisantz et al.has used a missile game where uncertainty was encoded with opacity to observe people’s ability to successfully shoot missiles and avoid harmless objects like birds and planes \cite{bisantz2011}. A number of other researchers have used similar games and simulations \cite{bisantz2011,kale2019,correll2018}. While there is value in knowing which visualization will lead to which outcome, the findings are limited to the task at hand and fail to provide a comprehensive understanding of decision-making processes.

In the medical field, researchers have investigated the impact of visualization design on gist reasoning. Feldman et al. compared the performance of 6 different visualization formats in inducing gist or detail-level processing ~\cite{feldman2007} (see figure 2). The results suggest that systematic ovals are likely the format that represents the best compromise for accurate processing of both gist and detailed information while also demanding relatively little effort. Hawley et al. found that viewing a pictograph was associated with adequate levels of both gist and verbatim knowledge, and that superior medical treatment choices were made in both cases~\cite{hawley2008}. 

Both Feldman et al. and Hawley et al. observed gist and verbatim knowledge when comparing two options: size of proportion and effectiveness of treatment respectively. There is a need to further investigate the decoding and extraction of both high level and detail-level information in other types of tasks such as Bayesian Reasoning. 





\section{Future Directions and Conclusion}

\textit{
\newline
``We live most of our lives guided by the impressions of System 1."
}
\vspace{-1em}
\begin{flushright}- Tversky and Kahneman \end{flushright}



To deepen our understanding of the effect of visualization on decisions, we need to look at \textbf{how} people make decisions. We suggest the following research questions, many of which are core to understanding the role of visualization.

\begin{itemize}
    \item Do people default to gist or verbatim reasoning when using visualization?
    \item Can a visualization design elicit gist or verbatim reasoning strategy? 
    \item How does gist and verbatim reasoning with visualization influence the decisions people make?
    
\end{itemize}



It is important to understand how people make decisions from visualizations. More specifically, understanding whether a visual encoding facilitates gist or verbatim reasoning can have huge implications for visualization designers. By expanding our knowledge in this area, we can tailor visualizations to our audience, or to a specific problem area. Bridging the gap between how psychologists and visualization researchers reason about decision-making can revolutionize the way we evaluate and design our visualizations.

\acknowledgments{This work was supported in part by the National Science Foundation under Grant No. 1755734.}

\bibliographystyle{abbrv-doi}

\bibliography{template}
\end{document}